\documentclass[prl,aps,twocolumn,showpacs,raggedbottom,noeprint,superscriptaddress,longbibliography,noeprint,10pt]{revtex4-1} %
\usepackage[T1]{fontenc}
\usepackage[latin9]{inputenc}
\usepackage{amsmath}
\usepackage{amssymb}
\usepackage{esint}
\usepackage{braket}
\usepackage{ulem}

\usepackage{graphicx}
\usepackage{hyperref}
\hypersetup{
	colorlinks=true,
	linkcolor=blue,
	filecolor=blue,
	citecolor = black,      
	urlcolor=cyan,
}

\usepackage[inline]{trackchanges}
\addeditor{Marcin}
\addeditor{Alexey}
\addeditor{PB}

\newcommand{\gsim}{{\;\raise0.3ex\hbox{$>$\kern-0.75em\raise-1.1ex\hbox{$\sim$}}\;}}

\usepackage{amsmath,epsfig,color,amssymb}
\usepackage{graphicx}
\usepackage{dcolumn}%
\usepackage{bm}%
\usepackage{graphicx}
\usepackage{amsmath, amsfonts, amssymb,mathrsfs}
\usepackage{pstricks}
\usepackage{amsxtra}
\usepackage{amsthm}
\usepackage{natbib}

\usepackage{color}
\usepackage{braket}
\usepackage{empheq}
\usepackage{empheq}
\usepackage[theorems,skins]{tcolorbox}
\usepackage{xspace} %

\definecolor{new}{rgb}{.08,.05,.8}
\newcommand{\new}[1]{{#1}}

\renewcommand{\remove}[1]{[REMOVE:\textcolor{gray}{#1}]}
\renewcommand{\remove}[1]{}

\def\metalearning{meta-learning\xspace}
\def\metaoptimizer{meta-optimizer\xspace}
\def\Metalearning{Meta-learning\xspace}

\def\ba{\begin{eqnarray}}
\def\ea{\end{eqnarray}}
\def\beq{\begin{equation}}
\def\eeq{\end{equation}}

\def\be{\begin{equation}}
\def\ee{\end{equation}}
\def\bea{\begin{eqnarray}}
\def\eea{\end{eqnarray}}

\def\bphi{\text{\boldmath$\phi\xspace$}}
\def\btheta{\text{\boldmath$\theta\xspace$}}%

\def\be{\begin{equation}}      %
\def\ee{\end{equation}}
\def\beu{\begin{equation*}}   %
\def\eeu{\end{equation*}}

\renewcommand{\paragraph}[1]{{\it #1.---}}

\def\calL{\mathcal{L}}

\usepackage{catchfilebetweentags}
\newif\ifsupp
\supptrue

\begin{document}
\title{Meta Hamiltonian Learning}
\date{\today}
\author{Przemyslaw Bienias}
\thanks{These authors contributed equally.}
\affiliation{Joint Quantum Institute, NIST/University of Maryland, College Park, Maryland 20742 USA}
\affiliation{Joint Center for Quantum Information and Computer Science, NIST/University of Maryland, College Park, Maryland 20742 USA}
\author{Alireza Seif}
\thanks{These authors contributed equally.}

\affiliation{Pritzker School of Molecular Engineering, University of Chicago, Chicago, Illinois 60637, USA}
\author{Mohammad Hafezi }
 \affiliation{Joint Quantum Institute, NIST/University of Maryland, College Park, Maryland 20742 USA}
\affiliation{Department of Electrical and Computer Engineering and Institute for Research in Electronics and Applied Physics, University of Maryland, College Park, Maryland 20742, USA}
\begin{abstract}
Efficient characterization of quantum devices is a significant challenge critical for the development of large scale quantum computers. We consider an experimentally motivated situation, in which we have a decent estimate of the Hamiltonian, and its parameters need to be characterized and  fine-tuned frequently to combat drifting experimental variables. We use a machine learning technique known as meta-learning to learn a more efficient optimizer for this task.  We consider training with the nearest-neighbor Ising model and study the trained model's generalizability to other Hamiltonian models and larger system sizes.  We observe that the meta-optimizer outperforms other optimization methods in average loss over test samples. This advantage follows from the meta-optimizer being less likely to get stuck in local minima, which highly skews the distribution of the final loss of the other optimizers. 
In general, meta-learning decreases the number of calls to the experiment and reduces the needed classical computational resources.
\end{abstract}
\maketitle

\paragraph{Introduction}%
Recently, there has been significant progress in experimental realizations of quantum computers~\cite{Debnath2016,wright2019benchmarking,arute2019quantum} and quantum simulators~\cite{Zhang2017}. This progress further motivates the need for characterization and validation of quantum devices, which is crucial for their precise control and operation. Resources required for complete characterization of a quantum state or a quantum process, known as quantum tomography, scales exponentially with the system size%
~\cite{Eisert2020a}. Specifically, learning the Hamiltonian of a quantum system is one of the extensively studied directions in the field of quantum characterization~\cite{Eisert2020a,carrasco2021theoretical}. \new{There are different approaches to address this problem more efficiently}: linear inversion based tomographic methods and compressed sensing~\cite{DaSilva2011,Shabani2011,Rudinger2015,Bairey2019}, maximum likelihood optimization or Bayesian methods~\cite{Granade2012,Wiebe2014,wang2017experimental,de2016estimation, Krastanov2019,Evans2019}, system identification algorithms \cite{Zhang2014b,Zhang2015,Sone2017,Wang2018f}, and techniques that exploit physical knowledge about the system to design model specific measurements to learn the Hamiltonian parameters \cite{Burgarth2009,Burgarth2011,DiFranco2009,Wang2015a,Qi2017}.
In addition to the above techniques, machine learning methods have been used to characterize quantum states and quantum processes~\cite{torlai2018neural,carrasquilla2019reconstructing,torlai2020quantum}. 

\new{Moreover, in noisy intermediate-scale quantum devices, control is limited, and it is not possible to prepare and measure many different configurations as required in many Hamiltonian estimation techniques. Current approaches make  use of time-traces of observables with a few easy-to-prepare initial states to estimate parameters of a model Hamiltonian\new{ by optimizing a cost function that quantifies how well a model's predictions agree with the observed measurement outcomes~\cite{Krastanov2019}}. Unfortunately, this fitting procedure requires a classical simulation of the quantum dynamics that is inherently difficult.}
Therefore, it is interesting to investigate whether machine learning tools can help with accelerating this process. 
\new{Here,} we use a machine learning technique known as \metalearning (also named as \textit{learning to learn})~\cite{Andrychowicz2016}, to design an algorithm that learns how to more efficiently optimize the cost function. This is achieved through  training the \metaoptimizer by solving samples of the optimization problem of interest. 

\begin{figure}[h!]%
\includegraphics[width= 1.00\columnwidth]{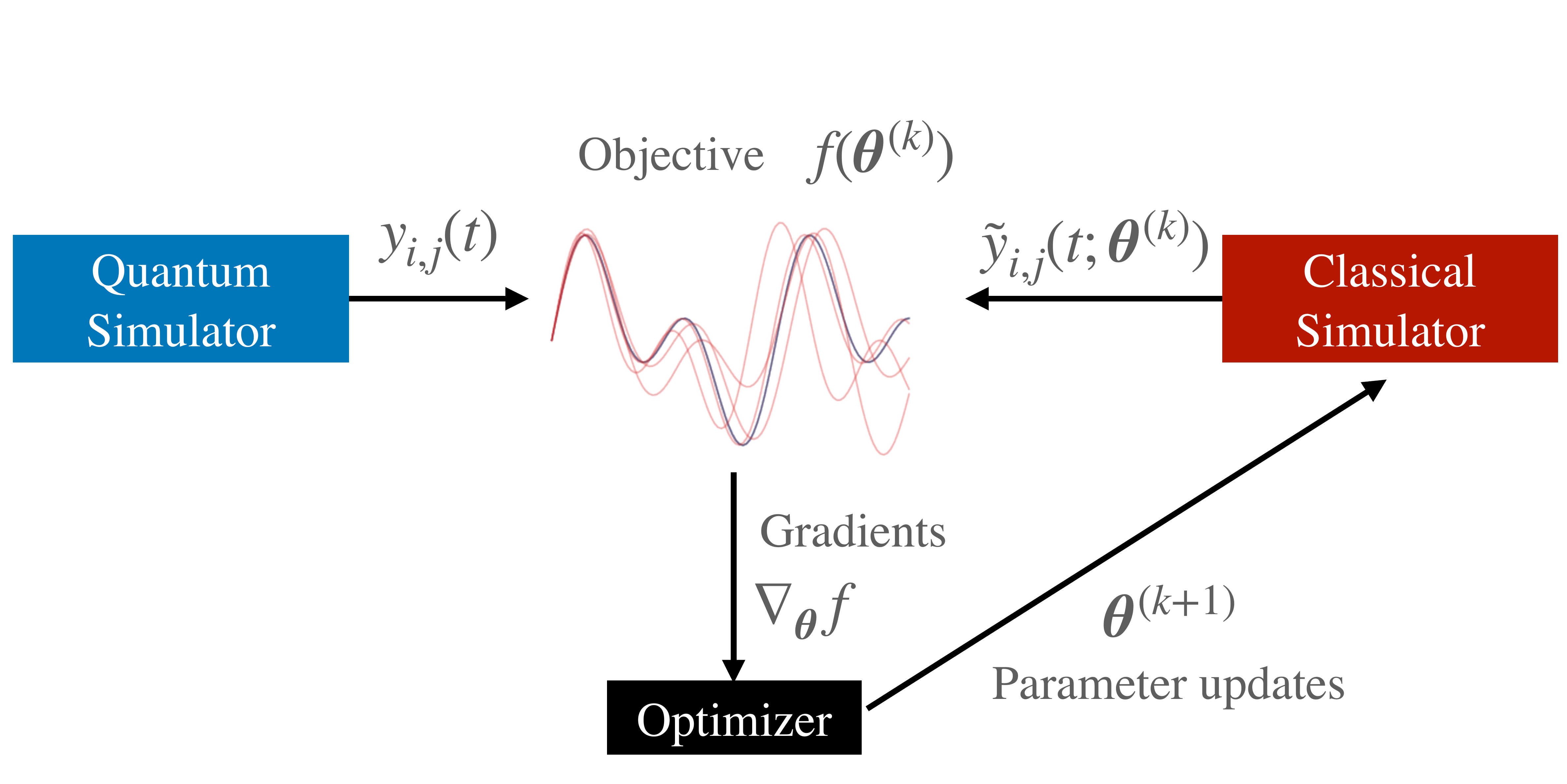}
\caption{
\textit{Schematic representation of the methods:}
The measurement signal $y_{i,j}(t)$ from the quantum device (blue) and our predicted signal $\tilde{y}_{i,j}(t;\btheta^{(k)})$ given the current estimate of the model parameters $\btheta^{(k)}$ (red) are compared to calculate the objective function $f(\btheta^{(k)})$. The gradient of the objective function, $\nabla_\btheta f$, is then provided to the optimizer, which updates our estimate for the parameters $\btheta^{(k+1)}$. This process is iterated to move model's prediction closer to the experimental observation.
}
\label{schematic}
\end{figure}

\begin{figure}[h!]%
\includegraphics[width= 1.00\columnwidth]{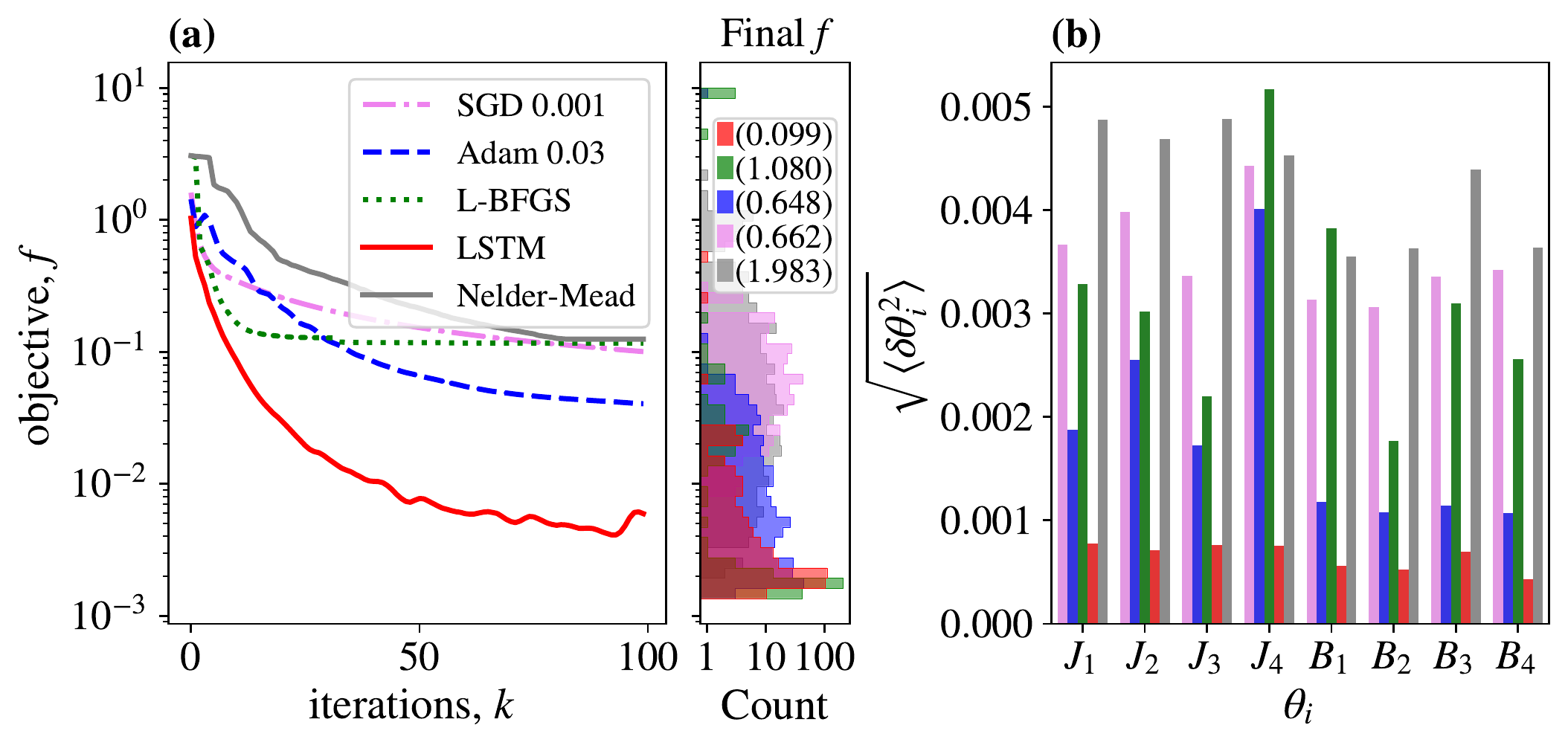}
\caption{
\textit{Performance of the \metaoptimizer}:
We test our optimizer on the same model as it was trained on, that is, on $N=4$ TFIM
 with stochastic noise $\sigma=0.001$. %
 (a) (left panel) The loss function $f$ versus the number of  iterations for different optimizers. (right panel) The histogram illustrating the loss values after the last iteration. The L-BFGS distribution with multiple instances of $f>10^{-2}$ corresponds to the optimizer being stuck in a local minima. Numbers in paranthesis indicate the standard deviation of the final $f$ value.
(b) The mean deviation $\delta\theta_i^2\equiv |\theta_i-\theta^*_i|^2$ from the true value $\theta^*_i$ of the Hamiltonian's parameters  after the last iteration. 
All results in these and the following figures are averaged over 300 epochs. 
}
\label{fig:fig2}
\end{figure}

\begin{figure}[h!]%
\includegraphics[width= 1\columnwidth]{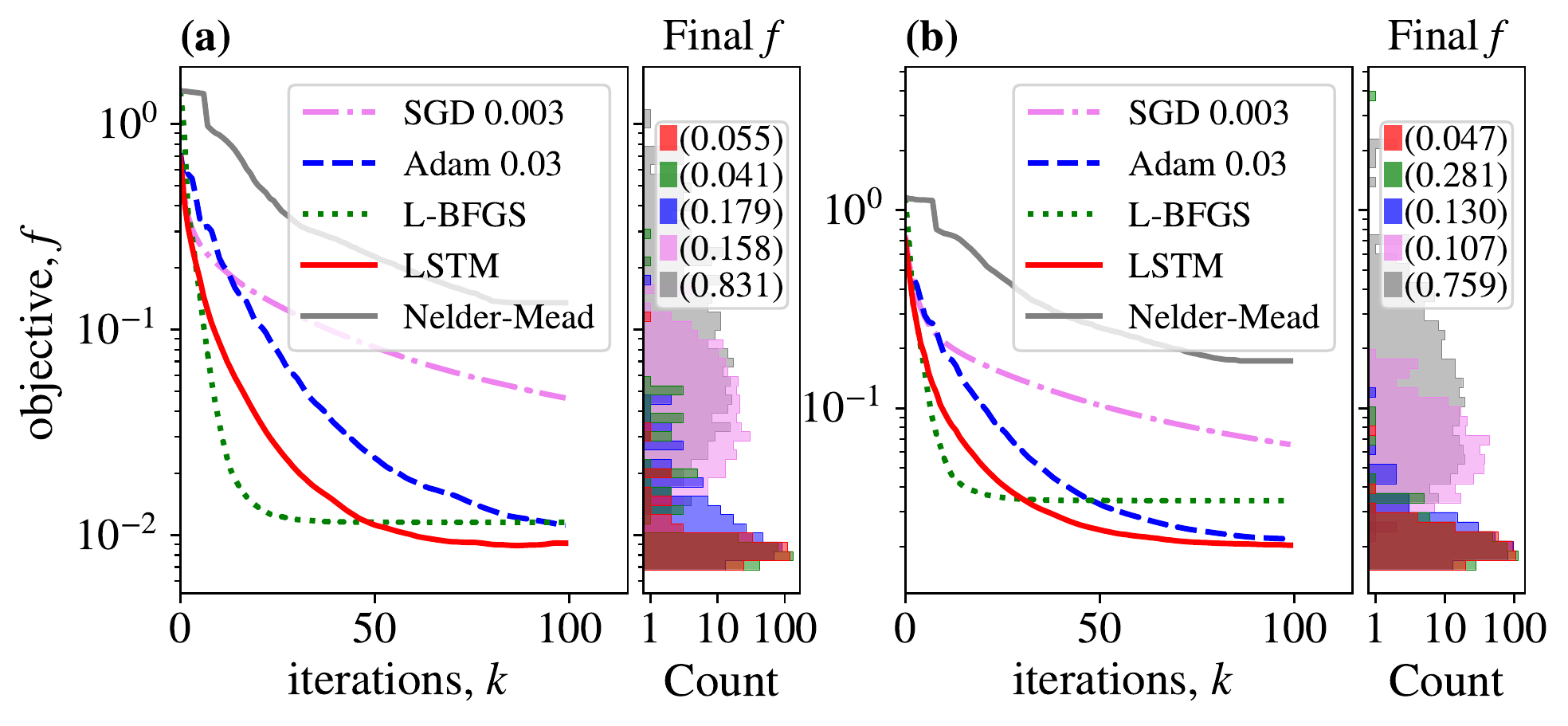}
\caption{
\textit{Generalizability to larger systems}: We test a model trained on $N=4$ qubits with the TFIM Hamiltonian \eqref{eq:transising} on larger systems with the same Hamiltonian and (a) 6 and (b) 7 qubits. The right side of each panel shows the spread of the final loss $f$ for different optimization schemes, \new{with the number in parenthesis indicating the corresponding standard deviations.} All other parameters are the same as those in Fig.~\ref{fig:fig2}. 
}
\label{fig:fig3}
\end{figure}

\begin{figure}[h!]%
\includegraphics[width= 1\columnwidth]{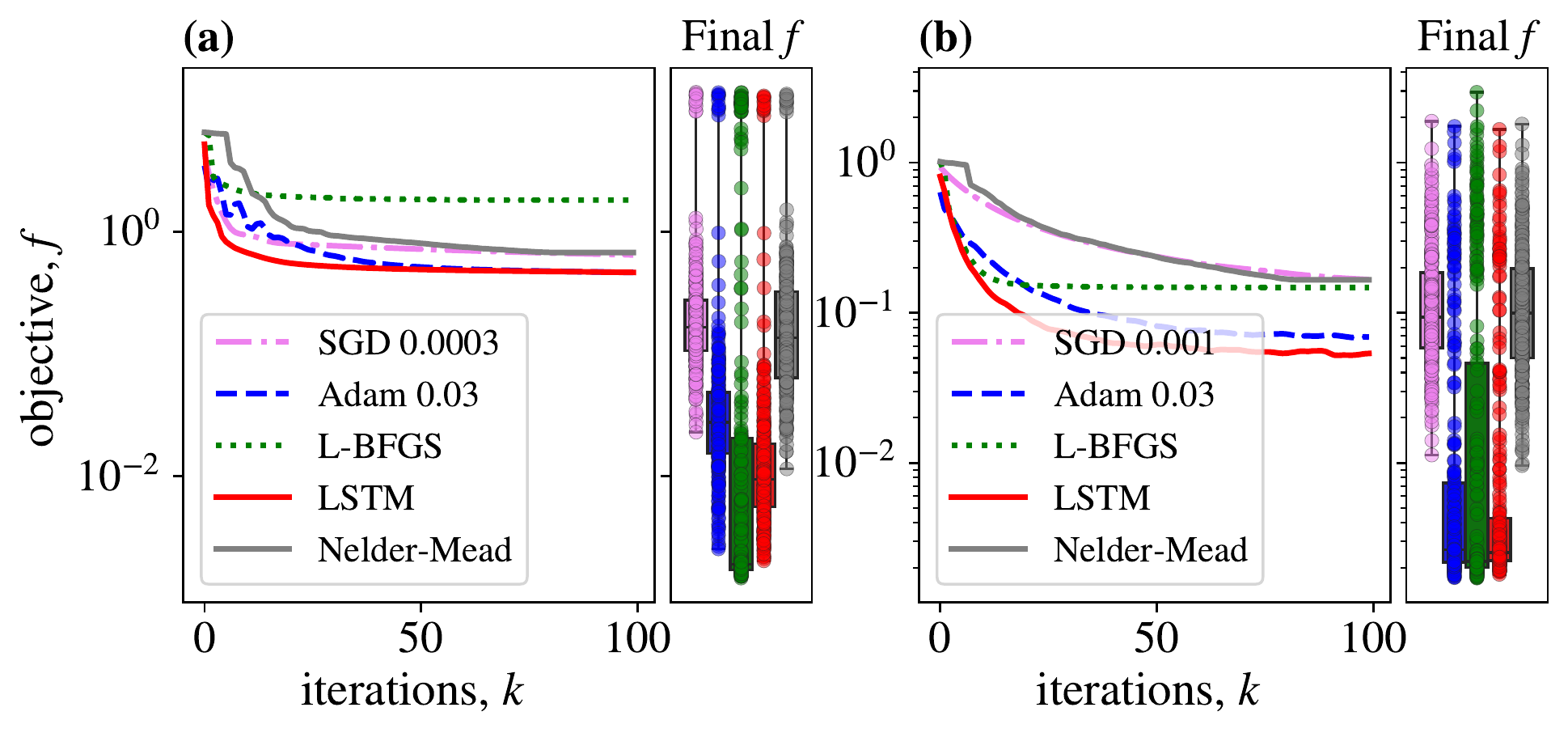}
\caption{
\textit{Generalizability to other models}: %
The model trained on $4$-qubit TFIM is tested on
(a) all-to-all Ising and (b) XY model.
{All other parameters are the same as those in Fig.~\ref{fig:fig2}.}  \new{The panels show the statistics of the final $f$ values.}%
We see better results for the XY model than for all-to-all due to the similar connectivity of the XY model and the TFIM.
}
\label{fig:fig4}
\end{figure}

\begin{figure}[h!]%
\includegraphics[width= 1\columnwidth]{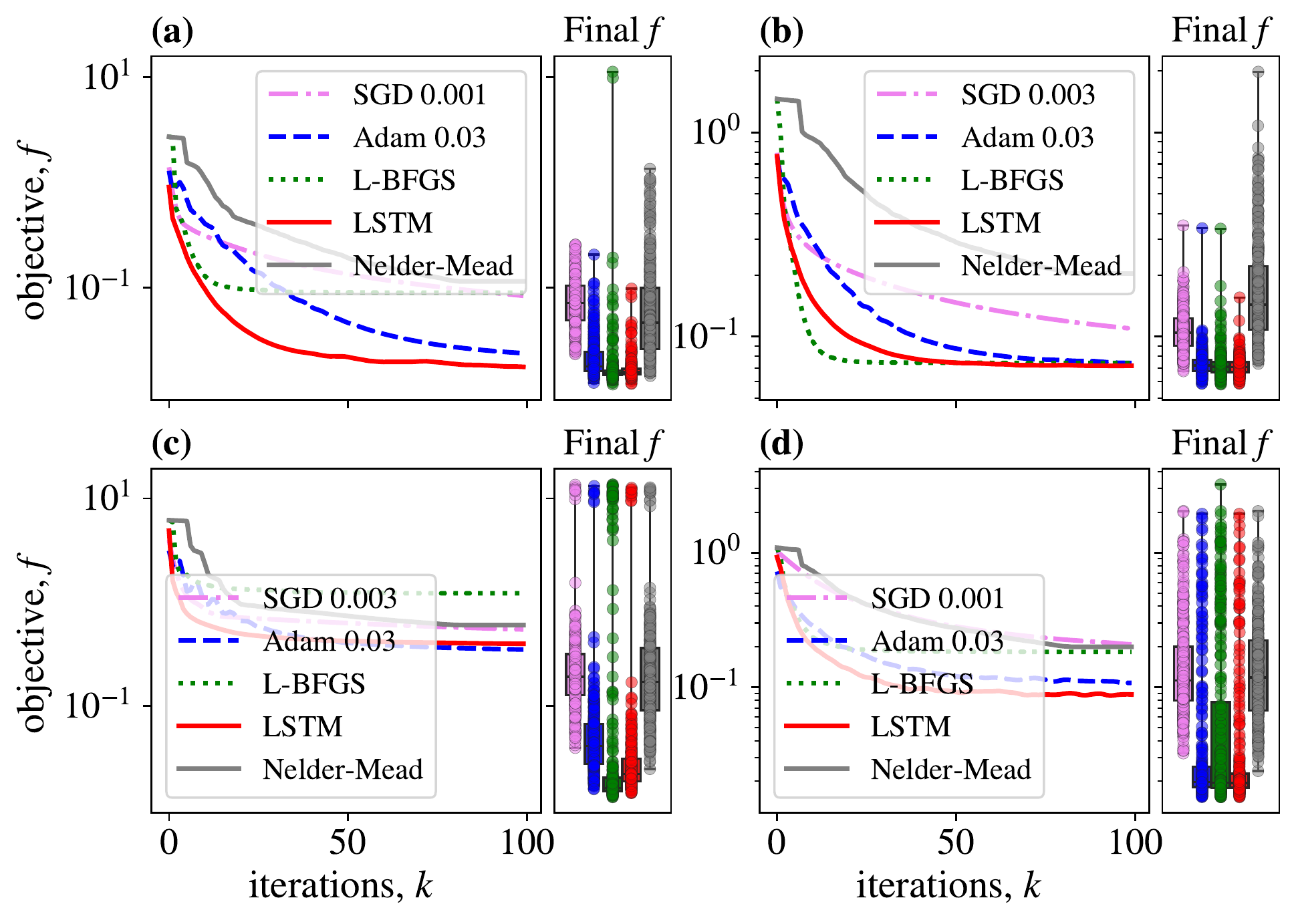}
\caption{
\textit{Generalizability to larger noise}: We study ten times larger stochastic noise $\sigma=0.003$ than for training and previous testing. (a)  TFIM with $N=4$, (b) TFIM for $N=6$, (c) all-to-all Ising model with $N=4$, and (d) XY model for $N=4 $.
}
\label{fig:fig5}
\end{figure}

Remarkably, we find that the \metaoptimizer algorithm (Fig.~\ref{schematic}) \new{performs better than the other optimizers we considered~(Fig.~\ref{fig:fig2}). Specifically, it achieves the lowest mean loss (0.006), a 6-fold decrease compared to the next best optimizer (0.040), over 300 test instances. Equivalently, the \metaoptimizer achieves the same loss as the next best optimizer with fewer number of iterations (16 versus 100).} Moreover, the spread of the test loss values characterized by the standard deviation is an order of magnitude smaller for the \metaoptimizer (0.099) compared to L-BFGS (1.080). As the data is highly skewed, we consider more detailed statistics later in this Letter.  An important feature of our \metaoptimizer is its flexibility in the number of input variables. \new{We show that  the algorithm successfully generalizes to both larger systems within the same class of models but with more parameters (Fig.~\ref{fig:fig3}), and to different classes of models (Fig.~\ref{fig:fig4}). We also demonstrate that the \metaoptimizer is robust against noise in all cases (Fig. \ref{fig:fig5}). }

\paragraph{System}%
We consider the task of inferring the Hamiltonian of the system from time traces of some observables in the experiment. Specifically, we model the system's Hamiltonian with $H(\btheta) = \sum_i \theta_i P_i$, where $\btheta$ is a vector that contains the parameters of the model $\theta_i$'s, and $P_i$'s are operators in the system's Hilbert space. Note that in general, any Hamiltonian for a system of qubits can be written in this form if $P_i$'s form a complete basis of operators in that Hilbert space, e.g., the Pauli basis. However, this set can be much more restricted due to geometric considerations and physical constraints on the nature of the couplings. The system is first prepared in different initial states $\{\rho_j\}$. 
Then, the system evolves and the expectation value of a set of observables $\{O_i\}$ is measured at different times for each initial state, that is $y_{i,j}(t) = {\rm{tr}}(\rho_j(t) O_i)$. We then minimize the least-square loss 
\begin{equation}\label{eq:optimizeeloss}
f(\btheta)=\sum_{i,j,t}(y_{i,j}(t)-\tilde{y}_{i,j}(t;\btheta))^2,
\end{equation}
where $\tilde{y}_{i,j}(t;\theta) = {\rm{tr}}\{\exp[-iH(\btheta)t]\rho_j(0)\exp[iH(\btheta)t] O_i\}$ is our model's prediction for $y_{i,j}(t)$. The resulting optimum, $\hat{\btheta}={\rm{argmin}_\btheta} f(\btheta)$, is our estimate for the parameters of the model~\cite{SupInfo}. %

Obviously, solving such an optimization problem is complicated, and we may not be able to find global minima under general circumstances. However, we focus on the case of tuning parameters, and assume that we have a close guess of the value of the parameters. For example, in an ion-trap experiment~\cite{haffner2008quantum} based on independently-measured sideband frequencies, %
one has a good estimate of the coupling strengths in the effective Ising Hamiltonian that governs the system. We should also note that a similar gradient-based optimization approach to Hamiltonian estimation has been previously considered in Ref.~\cite{Krastanov2019} by measuring the system's response to a control Hamiltonian. In contrast, we consider the response of the system to multiple initial states, which might be simpler in an experiment.

\paragraph{\Metalearning}%
In the machine learning context, learned representations and features often result in a better performance than what can be obtained with hand-designed representations and features~\cite{Goodfellow2016Book}. %
Still, optimization algorithms themselves are mostly designed by hand.
However, an optimization algorithm's design can be formulated as  a learning problem, so-called meta-learned optimization.
It has been shown that meta-learned optimization outperforms generic hand-designed competitors in many tasks~\cite{Andrychowicz2016}.
The application of this concept to the quantum domain has been recently considered in the context of the variational quantum algorithms~\cite{Verdon2019,Wilson2019}. %

To understand the concept of \metalearning, it is worth comparing it to the conventional optimization algorithms. 
For example, consider gradient descent based optimizers for a differentiable cost function $f$. The parameters $\btheta$ are updated in each optimization step $k$ via 
$\btheta^{(k+1)} = \btheta^{(k)} - \alpha_k \nabla f(\btheta^{(k)})$, where $\alpha_k$ is the learning rate at step $k$. In the \metalearning approach, however, the idea is to
replace the hand-designed update rules with a learned update rule, $\mathbf{g}_k$, parameterized by $\bphi$, such that
\begin{equation}
  \btheta^{(k+1)} = \btheta^{(k)} + \mathbf{g}_k(\nabla f(\btheta^{(k)}), \bphi).
  \label{eq:meta-update}
\end{equation}
We explicitly used the subscript $k$ in $\mathbf{g}_k$, to indicate that the update rule can depend on the optimization step $k$. We follow Ref.~\cite{Andrychowicz2016}'s approach and model the updates as the output  of a recurrent neural network (RNN). The step dependence of the updates are implemented  through the dynamically updated hidden state of the RNN, $\mathbf{h}_k$. %
Therefore, $\mathbf{g}_k$ is explicitly given by $[\mathbf{g}_{k},\mathbf{h}_{k+1}] = m(\nabla f(\btheta^{(k)}),\mathbf{h}_k, \bphi)$, where $m$ is the recurrent neural network function---referred to as the \metaoptimizer---and $\bphi$ encodes the trainable parameters of the RNN. Specifically, we use a Long Short-Term Memory (LSTM) neural network, %
which is highly effective in processing sequential data. Compared to common RNNs, LSTM have the ability to forget and add new information as time runs, which solves the vanishing gradient problem encountered in training other RNNs~\cite{pascanu2013difficulty,SupInfo}. %
The performance of the optimizer is quantified with the expected loss, $\mathbb{E}_{f}[f(\hat{\btheta})]$, over a distribution of functions $f$. To train the \metaoptimizer we can directly optimize the expected loss. However, it has been shown~\cite{Andrychowicz2016} that for the purpose of training the \metaoptimizer, it is advantageous to  instead minimize
\begin{equation}
    \calL(\bphi)
    = \mathbb{E}_{f} \left[\sum_{k=1}^T  f(\btheta^{(k)})\right],
    \label{eq:meta-optimization}
\end{equation}
that is the expected loss over optimization trajectories with the total of $T$ steps. 
As noted earlier, to tune and characterize the Hamiltonian parameters of a quantum system, one approach is to minimize Eq.~\eqref{eq:optimizeeloss} given estimates of $y_{i,j}(t)$ obtained from the experiment. Such an optimization-based approach for estimating Hamiltonian parameters fits nicely in the \metalearning framework. Hence, we take advantage of automatic differentiation techniques that allows us to readily calculate $\nabla_{\bphi} \mathcal{L}$ and train a \metaoptimizer for Hamiltonian learning.  
\paragraph{Setup}%
The \metaoptimizer $m$ is first trained on a system described by the nearest-neighbor Ising model. We then study the trained model's generalizability under various circumstances, such as systems with more spins, different Hamiltonians, and larger noise in the input.

The training data is generated by numerically evaluating the quench dynamics in the nearest-neighbor transverse field Ising model (TFIM) with the Hamiltonian
\begin{equation}\label{eq:transising}
    H = \sum_{i}^N J_i X_i X_{i+1} + \sum_{i}^N B_i Z_i,
\end{equation}
consisting of $N=4$ qubits with periodic boundary conditions. We use the convention $\btheta=\{J_1,...,J_N,B_1,...,B_N\}$.
We consider two easy to prepare initial states $\rho_j(t=0)$ with $j\in\{X,Z\}$, corresponding to the state with all qubits aligned along either $X$ or $Z$ directions, respectively~\cite{SupInfo}. %
The training time series data $y_{i,j}(t)$ are the probabilities of detecting the system with the initial state $\rho_j(0)$ in the $i$th computational basis state $\ket{i}$ at different times $t$. We model experimental and statistical errors by adding noise to the input data. The added noise is drawn from a normal distribution with mean zero and variance $\sigma^2$. Moreover, to be consistent with the laws of physics, we constrain the obtained noisy data such that $0\leq y_{i,j}(t)\leq 1$.

To train the \metaoptimizer neural network, we approximate the expectation value in the loss function \eqref{eq:meta-optimization} by the average over finitely many samples of $f$. Each sample corresponds to a realization of the Hamiltonian \eqref{eq:transising}. In the spirit of the stochastic gradient descent (SGD) algorithm~\cite{Goodfellow2016Book}, we update the parameters $\bphi$ of the \metaoptimizer through the estimated gradient of the loss function~\eqref{eq:meta-optimization} from a single random realization of the Hamiltonian and its corresponding time series. This process is then iterated many times with different realizations of the Hamiltonian~\eqref{eq:transising}. The initial value of the $\btheta$ at the beginning of each epoch, $\btheta^{(0)}$, reflects our prior knowledge about the problem. We choose ${\btheta}^{(0)}\sim\mathcal{N}(\btheta^*,\sigma_{\rm{in}}^2 I)$, where $\mathcal{N}(\boldsymbol{\mu},\boldsymbol{\Sigma})$ is the multivariate normal distribution with mean $\boldsymbol{\mu}$ and covariance matrix $\boldsymbol{\Sigma}$, and $\btheta^*$ is the true value of the parameters of the Hamiltonian in that epoch. 
Throughout the paper, we set $\sigma_{\rm{in}}=0.1$, and the overall energy scale to 1. 
During training we consider $\sigma=0.001$, 50 equally spaced times $t\in [0,10]$, elements of $\btheta$ drawn identically at random from the uniform distribution  $\mathcal{U}(1,2)$. We train the \metaoptimizer over $10^4$ epochs, with $T=100$ steps in each epoch. We assess the performance of the \metaoptimizer every 100 epochs and at the end choose the best performing model. %
\paragraph{Results}%
After initial training, we first test the \metaoptimizer on the same class of Hamiltonian with the same number of qubits, but with newly sampled parameters. We compare the average values of the optimizee objective function $f$ (given by Eq.~\eqref{eq:optimizeeloss}) versus the number of optimization steps (called iterations) $k$ obtained from our \metaoptimizer and other well-known algorithms. 
The results, shown in Fig.~\ref{fig:fig2}, indicate that in most cases, the \metaoptimizer outperforms derivative-free Nelder-Mead method~\cite{Nelder1965}, and first-order optimization methods, such as Adam~\cite{Kingma2014} and SGD~\cite{Goodfellow2016Book} with fine-tuned learning rate~\cite{SupInfo}. %
It also performs better on average than L-BFGS~\cite{liu1989limited}, a second-order optimization method; however, the advantage, in this case, is less clear. 
To better understand the performance of our LSTM optimizer, we consider the statistics of the final value of the objective function $f$, in addition to its average over optimization steps. We observe that while the \metaoptimizer is less likely to get stuck in local minima, which leads to a lower average cost, L-BFGS attains a similar final value in cases that it does not get stuck, see histograms in Fig.~\ref{fig:fig2}(a). We also show  how a smaller value of the objective function translates to a better estimate of the Hamiltonian parameters in Fig.~\ref{fig:fig2}(b).

The flexibility of the input dimension~\cite{SupInfo} %
of the \metaoptimizer allows us to assess its ability to generalize to models that are different than the one used in the training. We examine how the information learned from the four-spin Ising Hamiltonian transfers to more complicated situations. We first test the generalizability of the \metaoptimizer on the nearest-neighbor Ising model with more spins, see Fig.~\ref{fig:fig3}. {We observe that the comparative advantage of LSTM persists for $N=6$, however its performance deteriorates for $N=7$ and is comparable to Adam optimizer with fine-tuned learning rate.}
Moreover, we now study the generalizablity of the \metaoptimizer to different classes of models with the same number of qubits as the training. We consider the all-to-all Ising model,
\begin{equation}
    H = \sum_{i > j}^N J_{i,j} X_i X_{j} + \sum_{i}^N B_{i} Z_i ,
\end{equation}
 and the XY model,
\begin{equation}
    H =  \sum_i^N J^x_i X_i X_{i+1} + J^y_i Y_i Y_{i+1} + \sum_i^N Z_i .
\end{equation}
Compared to the TFIM, in addition to different quench dynamics, these models also have more Hamiltonian parameters  for a fixed system size.
In Fig.~\ref{fig:fig4}, we see that %
LSTM still outperforms SGD and Adam in the mean. %
\new{The performance of all of the optimizers is degraded compared with TFIM, resulting from an enhanced probability of getting stuck in local minima. Therefore, we turn to box plots (Fig.~\ref{fig:fig4}) to better compare the optimizer's performance in the presence of highly skewed data. We observe that boxes, which indicate the  75th and 25th percentiles are close for Adam, L-BFGS, and LSTM, however, the mean of L-BFGS is much larger than the other two because of the mentioned outliers. Note that in these plots, the whiskers show the range of the values. }
%

Lastly, we study the robustness of the \metaoptimizer against increased noise in all cases considered previously.  In Fig.~\ref{fig:fig5} we show results for $\sigma=0.003$ being three times greater than training and testing studying up to now. 
{For all cases (except the slightly worsened performance in the all-to-all Ising model)} the LSTM is still the best performing optimizer and the results are qualitatively the same as those with smaller noise in Figs.~\ref{fig:fig2},~\ref{fig:fig3}, and~\ref{fig:fig4}. 

%

\paragraph{Discussion}%
The trained \metaoptimizer in this work improves the classical processing resources required for calibrating experiments. Specifically, it reduces the amount of time needed to run expensive classical calculations to evaluate the cost function and its derivatives compared to other first-order optimization methods. The optimizer is also flexible and generalizes to various models. We believe that our approach is most useful when the \metaoptimizer is embedded in the daily calibration routine of an experiment, and is trained and tested on the same Hamiltonian model in the presence of considerable measurement noise. 

While we have shown that our method performs well on systems larger than what it is trained on, it is ultimately limited by the classical simulability of the system of interest. In both of the training and testing stages, we used exact diagonalization to evaluate the exact derivatives of the cost function. To push the limits of the applicability of our method, it is possible to train on smaller systems where exact diagonalization and automatic differentiation are available, and use efficient numerical simulation techniques such as those based on Matrix Product States (MPS) and Operators (MPO)~\cite{schollwock2011density} combined with numerical estimations of the derivatives during the test time. Another possibility is to replace the classical simulation in the test time with a well-calibrated quantum system~\cite{Wiebe2014} and use the output time series to estimate the gradient of the cost function numerically. 

Moreover, in this work, we focused on a gradient-based technique for estimating the Hamiltonian parameters of a physical system. It is also possible to use derivative-free methods, such as Bayesian optimization for the same task~\cite{Evans2019}. \Metalearning can again be used in this context to develop task-specific optimizers~\cite{Chen2016,Verdon2019}. In this approach, it is only necessary to have gradient information during the training. Therefore, in the Hamiltonian estimation problem, it is possible to train the \metaoptimizer on a small system where gradient information is readily available, and test on larger systems. The cost function, which still needs to be provided at every step in the test time, can again be evaluated using another quantum system or MPS based methods. 
\new{Additionally, it might be advantageous to consider alternative \metalearning techniques to the LSTM-based approach~\cite{li2016learning,li2017learning} in the context of Hamiltonian learning.}

%

Finally, while we studied %
parameter estimation in a Hamiltonian system, it is also possible to use the techniques developed here to study open quantum systems and Lindblad learning. In this case, automatic differentiation conveniently provides a gradient of the cost function with respect to the Lindblad parameters as shown in Ref.~\cite{Krastanov2019}, which is the main ingredient for training the \metaoptimizer.

\acknowledgments{
	{\it Acknowledgments.---}
	We thank Patrick Becker, Norbert M. Linke, Paraj Titum, Jiehang Zhang for insightful discussion. 
AS and MH acknowledge support by the Physics Frontier Center at JQI, ARO-MURI and the U.S. Department of Energy, Office of Science, Quantum Systems Accelerator (QSA) program. PB acknowledges funding by DoE ASCR Accelerated Research in Quantum Computing program (award No. DE-SC0020312), U.S. Department of Energy Award No. DE-SC0019449, the DoE ASCR Quantum Testbed Pathfinder program (award No. DE-SC0019040), NSF PFCQC program, AFOSR, AFOSR MURI, and ARO MURI.

This work was completed before AS joined the University of Chicago.}

%
%
%
 \bibliography{mainNotes.bib,library.bib,mainNotes_alireza_pb.bib,additional.bib}
%
%
\clearpage
\begin{widetext}
\begin{center}
{\Large \centering Supplemental material}
\end{center}
\setcounter{figure}{0}

\makeatletter

\setcounter{equation}{0}
\setcounter{figure}{0}
\setcounter{table}{0}

\renewcommand{\thefigure}{S\@arabic\c@figure}
\renewcommand{\thesection}{S.\Roman{section}}
\renewcommand \theequation{S\@arabic\c@equation}
\renewcommand \thetable{S\@arabic\c@table}

In this supplement, we present technical details.
For details of our LSTM construction and the effect of the size of the LSTM on the results see Sections~\ref{app:lstmdetails}~and~\ref{app:hiddennumber}.
In Sec.~\ref{app:otherlrs}, we discuss choosing the appropriate learning rate for the Adam and SGD optimizers.
In Sec.~\ref{app:alternateloss},  we  present a discussion on using a different loss function.
In sec.~\ref{app:initialstates}, we present a discussion on the effect of the choice of initial states on the optimization.

\section{LSTM details}\label{app:lstmdetails}
We build the \metaoptimizer in our work using the coordinate-wise LSTM construction from Ref.~\cite{Andrychowicz2016}. 
As noted in the main text, the incremental update of variable $\btheta$ is given by $\mathbf{g}_k$, where $\mathbf{g}_k$ is explicitly given by $[\mathbf{g}_{k},\mathbf{h}_{k+1}] = m(\nabla f(\btheta^{(k)}),\mathbf{h}_k, \bphi)$.  
Here, $m$ is modeled by an LSTM. The coordinate-wise construction essentially means that each component $\theta_i$ has its own hidden state, but the weights of the network are shared between all the variables. 
Specifically, let $m'[(\nabla^{(k)})_i,\mathbf{h}_{k,i},\bphi)]$ denote an LSTM that act on a single coordinate $\theta_i$, where we used $\nabla^{(k)}$ to shorten the notation of $\nabla f(\btheta^{(k)})$. 
Therefore, $\theta_i^{(k+1)} = \theta_i^{(k)} + g_{k,i}$, where $[g_{k,i},\mathbf{h}_{k+1,i}] = m'((\nabla^{(k)})_i,\mathbf{h}_{k,i}, \bphi)$. We then construct the full LSTM by concatenating the LSTMs and their hidden states for each coordinate to get $\mathbf{g}_k = [g_{k,1},\dots,g_{k,n}]$ and $\mathbf{h}_k = [\mathbf{h}_{k,1},\dots,\mathbf{h}_{k,n}]$, where $n$ is the number of coordinates. This construction provides flexibility in input dimension, and is also invariant to the ordering of the variables~\cite{Andrychowicz2016}. To simplify the training, we also assume that at each step $\nabla f(\btheta^{(k)})$ is obtained externally, and is independent of the parameters $\bphi$ of $m$. 

\section{Different learning rates for Adam and SGD}\label{app:otherlrs}
In this section, we show results for Adam and SGD with different learning rate (see  Fig.~\ref{figSLearningRate}).
Based on these results we choose the optimal learning rate $\eta$ which is used for the comparison with L-BFGS, Nelder-Mead, and LSTM in the main text.
\begin{figure}[h]
\includegraphics[width= .8\columnwidth]{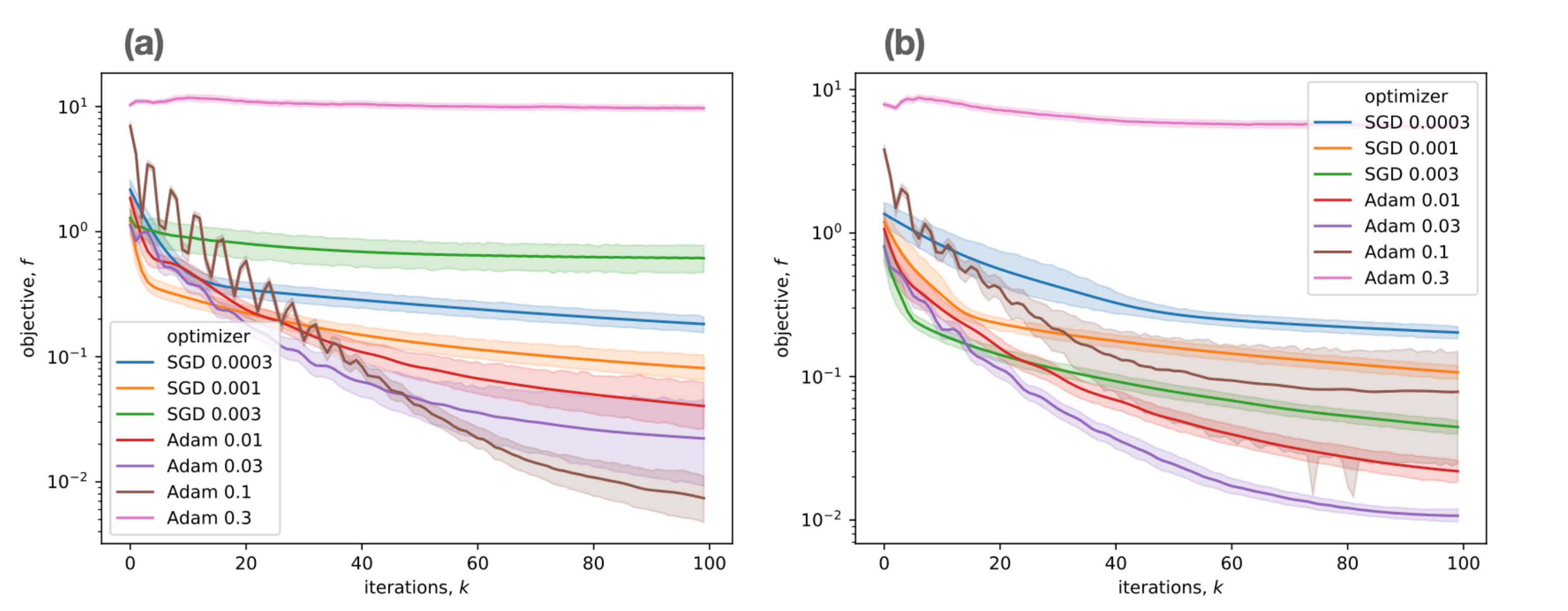}
\caption{
Results for transverse field Ising model
 with stochastic noise $\sigma=0.001$. 
The loss function $f$ versus the number of  iterations for different optimizers and different learning rates for (a) $N=4$ and (b) $N=6$. 
We see that optimal learning rates are: $\eta_{\text{Adam}}=0.03$ and  $\eta_{\text{SGD}}=0.001$ for $N=4$ and $\eta_{\text{Adam}}=0.03$ and  $\eta_{\text{SGD}}=0.003$ for $N=6$.
The shaded region is the 95\% confidence interval obtained by bootstrapping (resampling).
}
\label{figSLearningRate}
\end{figure}


\section{Kullback-Leibler divergence as a cost function}\label{app:alternateloss}
The observations we considered in this work, $y_{ij}(t)$, form a probability distribution over the basis states populations indexed by $i$ for a fixed $j$ and $t$. In this case, we can also use the Kullback-Leibler divergence to characterize the distance between the observed and model distributions and get 
\begin{equation}
\label{eq:optimizeelossKL}
f_{\rm{KL}}(\btheta)=\sum_{j,t} D_{\rm{KL}}[y_{:,j}(t)||\tilde{y}_{:,j}(t;\btheta)],
\end{equation}
where we used the notation $y_{:,j}(t)$ to indicate the discrete population distribution over all $i$s at a fixed realization $j$ and time $t$, and $D_{\rm{KL}}[P||Q] = \sum_i P_i \log(P_i/Q_i)$.
The test results (Fig.~\ref{figS_KL}) show qualitatively similar behavior to what we saw when using the squared loss: The performance of the LSTM for $N=4$ is superior to other optimizers, and the advantage shrinks when a $N=6$ system is considered.
\begin{figure}[h]
\includegraphics[width= .8\columnwidth]{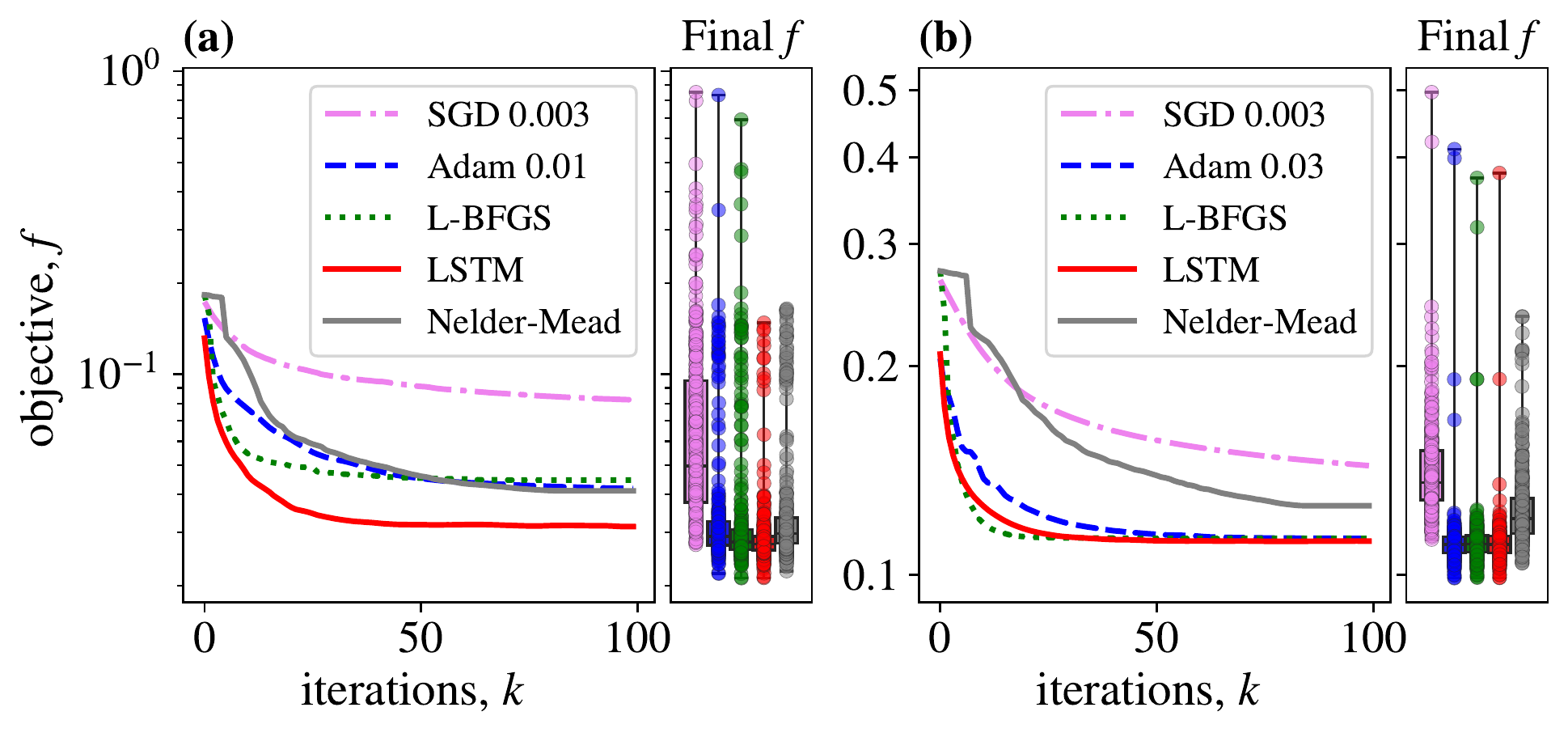}
\caption{
Results for transverse field Ising model
 with stochastic noise $\sigma=0.001$. 
The loss function proportional to the Kullback-Leibler divergence  versus the number of  iterations for different optimizer for $N=4$ and $N=6$, see (a) and (b), respectively.
}
\label{figS_KL}
\end{figure}

\section{Different initial states}\label{app:initialstates}
In the main text, we study results based on two easy to prepare initial states $\rho_j(t=0)$ with $j\in\{X,Z\}$, corresponding to the state with all qubits aligned along either $X$ or $Z$ directions.
In Fig.~\ref{figS_initState} we show the results for the training and testing based on initials state $\rho_X(t=0)$ and $\rho_Z(t=0)$, see Fig.~\ref{figS_initState}(a) and (b), respectively. 
{We see that for testing with the same system size, LSTM outperforms the other methods in the same sense that it did in} Fig.~(2) in the main text.
\begin{figure}[h]
\includegraphics[width= .8\columnwidth]{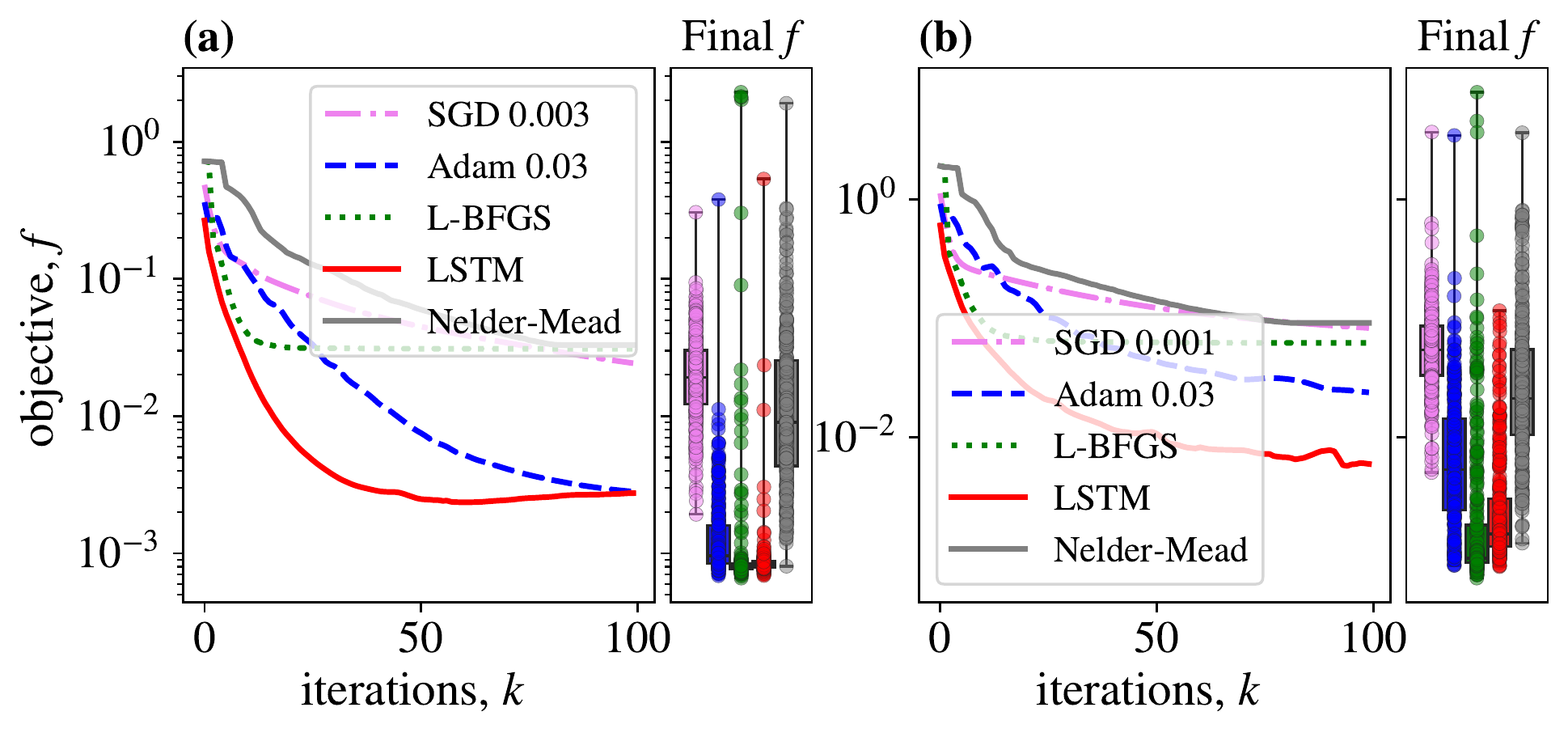}
\caption{
Results for transverse field Ising model
 with stochastic noise $\sigma=0.001$ and $N=4$. 
(a) in $X$ direction, (b) in $Z$ direction.
}
\label{figS_initState}
\end{figure}

\section{Different number of hidden neurons}\label{app:hiddennumber}
The internal structure of the meta-optimizer can be changed.
The LSTM cell is characterized by, e.g., the number of hidden neurons, $N_h$. All the results presented in this paper are for $N_h=20$, following Ref.~~\cite{Andrychowicz2016}.
In Fig.~\ref{figSHidden}(a) and (b), we show results for training and testing with $N_h=10$ and $N_h=30$, respectively. {We do not observe any qualitative difference between the two cases.} 
\begin{figure}[h]
\includegraphics[width= .8\columnwidth]{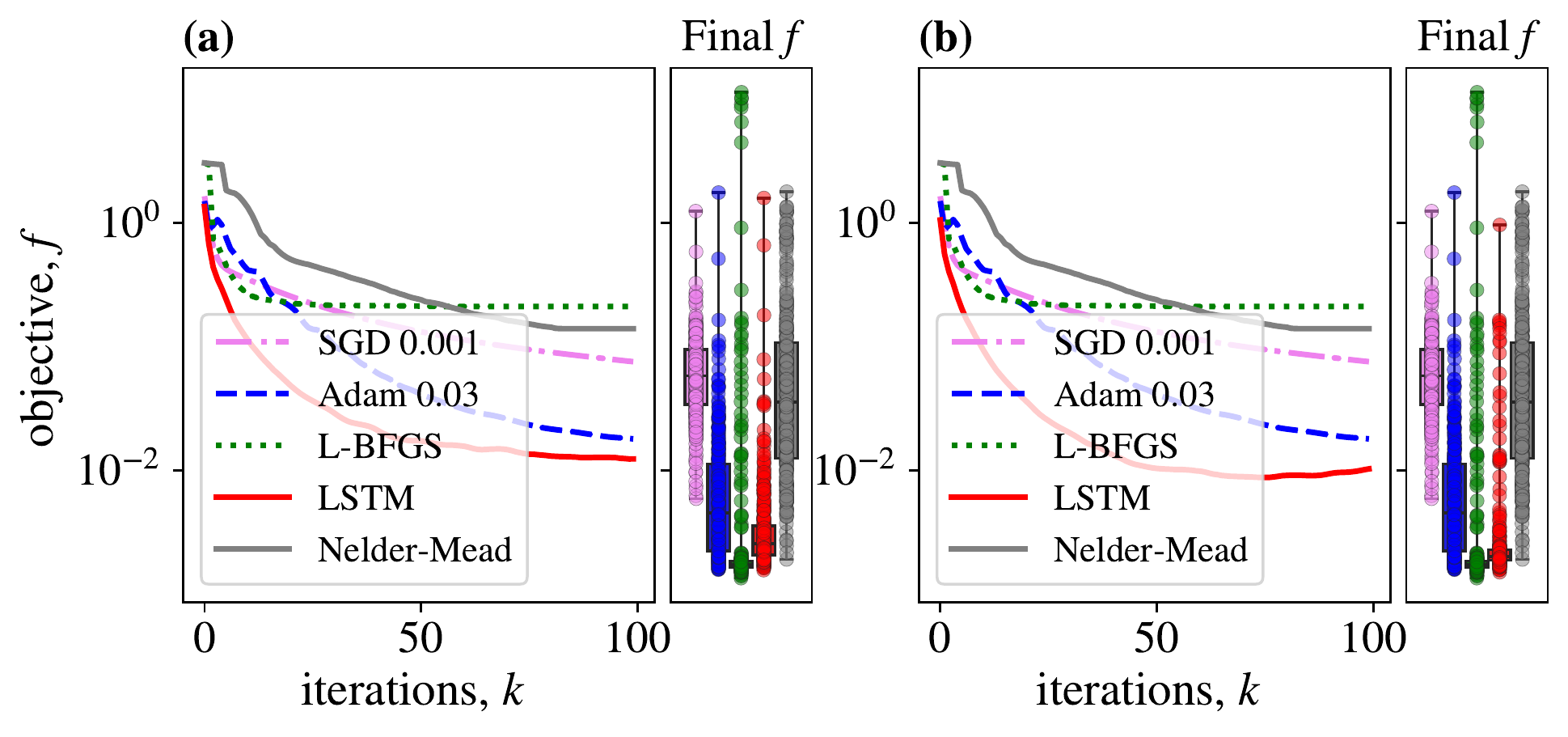}
\caption{
Results for transverse field Ising model with stochastic noise $\sigma=0.001$, $N=4$, with $N_h=10$ and $N_h=30$, see (a) and  (b), respectively.
}
\label{figSHidden}
\end{figure}

\clearpage
\end{widetext}
\end{document}

\ifsupp
\def\thesection{\Roman{section}}
\setcounter{secnumdepth}{2}
\widetext
\pagebreak
\ExecuteMetaData[supplement]{document}
\fi	
\end{document}